\documentclass[12pt,a4paper]{article}

\usepackage[utf8]{inputenc}
\usepackage{amssymb}
\usepackage{multicol}
\usepackage[margin=2.5cm]{geometry}
\usepackage{setspace}
\usepackage{graphicx}
\usepackage{url}
\usepackage{color}
\usepackage{hyperref}
\graphicspath{ {images/} }
\usepackage{tikz}
\usepackage{amsmath}
\usepackage{physics}
\usepackage{ulem}
\usepackage{xcolor}
\usepackage{centernot}
\usepackage{amsthm}
\newtheorem{definition}{Definition}

\newtheorem{proposition}{Proposition}
\newtheorem{principle}{Principle}
\usepackage{fancyhdr}

\usepackage{centernot}

\begin{document}

\doublespacing

\title{Quantum Logic and Meaning}

\author{Sebastian Horvat \\ Department of Philosophy \& Faculty of Physics \\ University of Vienna \\ \\ Iulian D. Toader \\  Institute Vienna Circle \\ University of Vienna }

\date{ }

\maketitle

\begin{abstract}

This paper gives a formulation of quantum logic in the abstract algebraic setting laid out by Dunn and Hardegree (2001). On this basis, it provides a comparative analysis of viable quantum logical bivalent semantics and their classical counterparts, thereby showing that the truth-functional status of classical and quantum connectives is not as different as usually thought. Then it points out that bivalent semantics for quantum logic -- compatible with realism about quantum mechanics -- can be maintained, albeit at the price of truth-functionality. Finally, the paper critically addresses Geoffrey Hellman's argument (1980) that this lack of truth-functionality entails a change of meaning between classical and quantum connectives.

\end{abstract}

\newpage 

\section{Introduction}


Recall that Birkhoff and von Neumann motivated their project ``to discover what logical structure one may hope to find in physical theories which, like quantum mechanics, do not conform to classical logic'' (Birkhoff and von Neumann 1936, 823) by their realization that the transition from classical physics to quantum mechanics (QM) requires that the commutativity of operators representing quantum observables be dropped. This transition, they argued, indicates that within the quantum domain classical logic (CL) must be replaced by a non-distributive quantum logic (QL). Putnam then argued that this QL is demanded by metaphysical realism, and more precisely by a realist interpretation of QM that does not postulate, say, hidden variables (Putnam 1968). He clearly emphasized that it is the \textit{classical} distributive law that must be dropped in QM. Arthur Fine responded that it is \textit{not} the classical distributive law that is dropped in QM - there is no clash between quantum non-distributivity and classical distributivity - so, as a consequence, the transition in physics does not require a revision of CL (Fine 1972). Geoffrey Hellman (1980) attempted to rigorously justify this intuitive response by proving that some QL-connectives are not truth-functional, which he took to support the view that the meaning of those connectives is different than that of their CL counterparts. Rather than a revision, QL is therefore presumably just a replacement of CL, exactly as Birkhoff and von Neumann appear to have intended it. 

The goal of our paper is to engage with this dialectic by investigating more thoroughly the formal-semantic similarities and differences between QL and CL, and by reflecting on how these bear upon Hellman's meaning-variance argument. The structure runs as follows. In section two, we first prepare the ground by formulating QL in an abstract algebraic framework. This will enable a rigorous and transparent comparison of some formal-semantic aspects of CL and QL. In section three, we proceed with a detailed analysis of viable bivalent semantics for QL, and we show that the truth-functional status of CL- and QL-connectives is not as different as commonly thought. In section four, we emphasize the seemingly forgotten fact that a bivalent semantics for QL, which may potentially support realism about QM, is indeed tenable, provided that one gives up the truth-functionality of some QL-connectives. In the same section, we also identify and discuss another formal condition that ought to be satisfied by any purportedly realist semantics. Finally, on this background, we come to address Hellman's argument in section five. We show that our formal-semantic analysis from the previous sections indicates that his argument can be improved, which brings us to propose a stronger argument, reinforcing and precisifying its conclusion that some QL-connectives have a different meaning than their CL counterparts. Our argument still depends on the same meaning-variance principle that was originally suggested by Hellman and, thus, on the use of truth-functionality as a constraint on synonymy, but also, perhaps surprisingly, on the way one formally explicates the validity of arguments with multiple premises. At the end of the paper, we point to some problems with the justification of these assumptions, but we defer a more extensive analysis -- one that goes beyond formal-semantic methods -- to future work.


\section{An Algebraic Formulation of Quantum Logic}
QL has been formulated in a variety of ways, both model- and proof-theoretically. Some of these formulations even yield different logics, in the sense that they validate different sets of arguments, thus making it more apt to speak of QLs, in the plural. Nevertheless, since the details and differences among these formulations are not relevant for our discussion, we will stick to the original one, given in terms of Hilbert lattices, albeit presented more abstractly with the help of the terminology and mathematical machinery employed by Dunn and Hardegree in their classic study of algebraic methods in logic (Dunn and Hardegree 2001). The main advantage of this presentation, as we will see shortly, is that the interpretation of sentences is not completely fixed, as in more standard presentations of QL (e.g., R\'edei 1998), thus ``erasing any intensional vestiges'' in the language of QL and enabling a clearer semantic comparison with CL.\footnote{Just to make clear, once again, our more abstract formulation defines the same QL as the one originally defined by Birkhoff and von Neumann in terms of Hilbert lattices.}
\subsection{Algebraic preliminaries}
Let $\mathcal{L}$ be a sentential language containing atomic sentences, conjunction `$\land$', disjunction `$\vee$' and negation `$\lnot$'. Let $S$ be the \textit{set of sentences} generated inductively from $\mathcal{L}$, in the usual way. We can straightforwardly define the \textit{algebra of sentences}.
\begin{definition}
The algebra of sentences $\mathcal{S}\equiv (S, O_{\land},O_{\vee},O_{\lnot})$ is an algebra whose operations on the carrier set $S$ are defined as follows: for any $a,b \in S$,
\begin{equation*}
    \begin{split}
        &O_{\land}(a,b)=a \land b\\
        &O_{\vee}(a,b)=a \vee b\\
        &O_{\lnot}(a)=\lnot a.
    \end{split}
\end{equation*}
\end{definition}
Next, we introduce the notion of a \textit{logical atlas}.
\begin{definition}
A logical atlas $\mathcal{A}$ is a pair $(\mathcal{P},\langle D_j \rangle)$, where $\mathcal{P}$ is a non-empty algebra and $\langle D_j \rangle$ is a family of proper subsets of the carrier set of $\mathcal{P}$. 
\end{definition}
We will refer to the sets $D_j$ as \textit{designated sets}. Logical atlases can be used to provide a semantics for our language. Indeed, an \textit{interpretation} of a language into an atlas is defined as follows:
\begin{definition}
Let $\mathcal{L}$ be a sentential language and $\mathcal{S}$ its associated algebra of sentences. Suppose $\mathcal{A}=(\mathcal{P},\langle D_j \rangle)$ is a logical atlas whose algebra $\mathcal{P}$ is of the same type as $\mathcal{S}$. An interpretation of $\mathcal{L}$ in $\mathcal{A}$ is any \textit{homomorphism} from $\mathcal{S}$ into $\mathcal{P}$.
\end{definition}
The elements of the carrier set of $\mathcal{P}$ can be thought of as \textit{propositions}, and each designated set $D_j$ can be thought of as a possible set of true propositions, i.e. a way ``the world could be''. An interpretation of a language into an atlas thus homomorphically assigns propositions to sentences, in accord with the principle of compositionality; since an atlas can contain multiple designated sets, an interpretation does not need to fully determine which propositions are true and which are not. We will call a triple $(\mathcal{L}, \mathcal{A}, I)$, where $I$ is the set of all interpretations of $\mathcal{L}$ in $\mathcal{A}$, an \textit{interpretationally constrained language}.

Let us now introduce the notion of \textit{valuation}.
\begin{definition}
Let $\mathcal{L}$ be a sentential language and $S$ its associated set of sentences. A valuation on $\mathcal{L}$ is any function from $S$ to the two-element set $\left\{\textbf{t},\textbf{f} \right\}$.
\end{definition}
Here we are taking the valuation of a language to be bivalent by definition: each sentence of the language is assigned either `true' or `false', thereby excluding third truth-values or truth-value ``gaps'' and ``gluts''. Also, there are no compositional restrictions on truth-value assignments to compound sentences, i.e. we are not assuming truth-functionality; more on this later.

The identification of designated sets with true propositions suggests a natural way of defining a class of valuations induced by an atlas.
\begin{definition}\label{atlas val}
Let $\mathcal{A}=(\mathcal{P},\langle D_j \rangle)$ be a logical atlas and $\mathcal{L_A}=(\mathcal{L}, \mathcal{A}, I)$ an interpretationally constrained language. For each interpretation $i \in I$, and index $j$, define a valuation $v^{(i)}_j$ as:
\begin{equation*}
    v^{(i)}_j(a)=\textbf{t}, \quad \text{if and only if} \quad i(a) \in D_j.
\end{equation*}
The class of valuations $C^*_{\mathcal{A}}$ induced by $\mathcal{A}$ is the set of all such valuations, i.e. 
\begin{equation*}
    C^*_{\mathcal{A}}=\left\{v^{(i)}_j, \quad \forall i \in I, \forall j\right\}.
\end{equation*}
\end{definition}
Now we are ready to introduce the notion of \textit{logical consequence} induced by an atlas. We say that a set of premises $\Gamma \subseteq S$ $\mathcal{A}$-implies a conclusion $a \in S$, or $\Gamma \vDash_{\mathcal{A}} a$, if the following holds:
\begin{equation}\label{atlas cons}
    \forall v \in C^*_{\mathcal{A}}: \quad \text{if} \quad \left(\forall \gamma \in \Gamma, v(\gamma)=\textbf{t} \right) \quad \text{then} \quad v(a)=\textbf{t}.
\end{equation}
Analogously, we say that a sentence is $\mathcal{A}$-valid, or $\vDash_{\mathcal{A}} a$, if $v(a)=\textbf{t}$ for all valuations $v$ in $C^*_{\mathcal{A}}$.

To summarize, an interpretationally constrained language $\mathcal{L_A}$ consists of a language endowed with a compositional semantics and contains sufficient structure -- namely, the family of designated sets pertaining to the logical atlas $\mathcal{A}$ -- to induce a class of valuations $C^*_{\mathcal{A}}$, thereby defining a logical consequence relation $\vDash_{\mathcal{A}}$ on the set of sentences. 

Let us present QL within this formalism.
 
\subsection{Quantum logic}
We will now use the definitions from the previous subsection in our presentation of QL. In particular, we will introduce \textit{quantum atlases} and their induced quantum-logical consequence relations.

Let $C(H)$ be the set of closed subspaces of a (potentially infinite-dimensional) Hilbert space $H$. We will say that the \textit{algebra associated to $H$} is the algebra $\mathcal{H}=(C(H),\cap,\sqcup, ^{\perp})$, where `$\cap$' and `$\sqcup$' are set-intersection and linear span, respectively, whereas the `$^{\perp}$' operation takes a subspace and outputs the corresponding orthogonal subspace. We define \textit{quantum atlases} $\mathcal{A_{H}}$ as follows:
\begin{definition}
A quantum atlas $\mathcal{A_{H}}=(\mathcal{H},\langle D_P \rangle)$ is an atlas where $\mathcal{H}=(C(H),\cap,\sqcup, ^{\perp})$ is the algebra associated to some Hilbert space $H$, and the family of designated subsets $\langle D_P \rangle$, whose index ranges over all 1-dimensional subspaces $P \in C(H)$, is defined as
\begin{equation*}
    \forall Q \in C(H): \quad Q \in D_P \quad \text{if and only if} \quad P \subseteq Q.
\end{equation*}
\end{definition}
A \textit{quantum interpretationally constrained language} $\mathcal{L_{A_H}}$ is then any triple $(\mathcal{L}, \mathcal{A_{H}}, I)$, where $\mathcal{L}$ is a sentential language, $\mathcal{A_{H}}$ is a quantum atlas and $I$ is the set of all interpretations of the language into $\mathcal{A_{H}}$. In particular, being homomorphic, the interpretations $i \in I$ satisfy
\begin{equation}
    \begin{split}
        &i(a \land b)=i(a)\cap i(b)\\
        &i(a \vee b)=i(a) \sqcup i(b)\\
        &i(\lnot a)=i(a)^{\perp},
    \end{split}
\end{equation}  
for all sentences $a,b$ in $S$. Following Definition \ref{atlas val} and Eq. \eqref{atlas cons}, we can associate to any atlas $\mathcal{A_H}$ an induced class of valuations $C^*_{\mathcal{A_H}}$ and an induced consequence relation $\vDash_{\mathcal{A_H}}$, which thereby lets us define the notion of quantum-logical consequence. In order to simplify the notation, we will henceforth write $\mathcal{C}^*_{H}$ and $\vDash_{H}$, instead of the more cumbersome $C^*_{\mathcal{A_H}}$ and $\vDash_{\mathcal{A_H}}$.
\begin{definition}
Let $\vDash$ be a consequence relation on a set of sentences $S$. We say that $\vDash$ is a quantum-logical (QL) consequence relation if there exists a quantum atlas $\mathcal{A_H}$, such that $\vDash$ and $\vDash_{H}$ coincide. 
\end{definition}
QL-consequence relations can also be given the following characterization, reminiscent of the one originally introduced by Birkhoff and von Neumann: for a Hilbert space $H$, a set of premises $\Gamma$ $H$-implies conclusion $a$ if for all homomorphisms $i$ from the algebra of sentences into the algebra associated to $H$, the following holds
\begin{equation}
    \bigcap_{\gamma \in \Gamma} i(\gamma)\subseteq i(a).
\end{equation}
Since the operations $\cap$ and $\sqcup$ on $C(H)$ are non-distributive, it follows that QL-consequence relations violate the law of distributivity, as we will now illustrate (see also Birkhoff and von Neumann 1936, 831).

Let $a_1,a_2,b \in S$ be atomic sentences. Consider an interpretation $i \in I$ that maps $a_1$ and $a_2$ into two mutually orthogonal 1-dimensional subspaces of $H$ that contain respectively vectors $\psi_1$ and $\psi_2$, and $b$ into the 1-dimensional subspace that contains vector $\frac{1}{\sqrt{2}}\left(\psi_1+\psi_2\right)$. It follows that $i(b \land (a_1 \vee a_2))=i(b)$ and $i( (b \land a_1) \vee (b \land a_2))=\mathbf{0}$, which implies $i(b \land (a_1 \vee a_2)) \nsubseteq i( (b \land a_1) \vee (b \land a_2))$, and thus $b \land (a_1 \vee a_2) \centernot\vDash_{H} (b \land a_1) \vee (b \land a_2)$. This failure of distributivity in QL is a direct consequence of the fact that experimental propositions in QM form a non-distributive lattice, which can be arguably understood as a consequence of the algebra of quantum-mechanical observables being non-commutative.

Note that we have defined the QL-consequence relations $\vDash_{H}$ relative to Hilbert spaces $H$: there may thus in principle be as many different consequence relations as there are (non-isomorphic) Hilbert spaces. Consider for a moment the parallel situation in classical logic (CL). Instead of defining the CL-consequence relation in terms of standard truth-tables or via some proof-theoretic axiomatization, one may introduce it by using atlases built on arbitrary Boolean algebras. Since there are many non-isomorphic Boolean algebras, one needs to prove that they all lead to one and the same logical consequence relation. Indeed, the latter turns out to be the case, due to the theorem that states that an equation holds in an arbitrary Boolean algebra if and only if it holds in the two-element Boolean algebra: thus, all these different Boolean algebras yield one and the same consequence relation, which can be defined using the standard truth-tables.\footnote{In algebraic terms, to say that an equation holds in an arbitrary Boolean algebra if and only if it holds in the two-element Boolean algebra means to say that the variety of Boolean algebras is generated by the two-element Boolean algebra, which can be seen as a consequence of the Stone representation theorem for Boolean algebras (Stone 1936).} On the other hand, there is no parallel result that holds for algebras associated to Hilbert spaces: for instance, as noted already by Birkhoff and von Neumann (1936, 832), modularity holds only in finite-dimensional Hilbert spaces (see also R\'edei 1998). There is thus, as suspected, more than one QL-consequence relation depending on the Hilbert space on which the semantics is built, which justifies keeping the index `$H$' in `$\vDash_{H}$'. 

\section{On Valuations in Quantum Logic}\label{formal}
In the previous section we introduced quantum atlases $\mathcal{A_H}$, their induced classes of valuations $\mathcal{C}^*_{H}$ and the induced relations of logical consequence $\vDash_{H}$. Notice that $\vDash_{H}$ are abstract relations on the set of sentences $S$, so despite being explicitly constructed with the help of classes $\mathcal{C}^*_{H}$ induced by quantum atlases, they are in principle definable by other means, e.g., by some proof-theoretic axiomatization, or by different classes of valuations. In this section we want to focus on the latter, i.e., we want to analyze some features of the collection of viable classes of valuations that respect $\vDash_{H}$. With this aim in mind, let us introduce the notion of \textit{$H$-class}.
\begin{definition}\label{val ql}
Let $\mathcal{L}$ be a sentential language and $S$ its set of sentences. Suppose the consequence relation $\vDash_{H}$ on $S$ is induced by a quantum atlas $\mathcal{A_H}$, for some Hilbert space $H$. An $H$-class $\mathcal{C}$ is any class of valuations on $\mathcal{L}$ that obeys the following. For all $\Gamma \cup \left\{a\right\} \subseteq S$:
\begin{equation}\label{valuations}
    \Gamma \vDash_{H} a \quad \quad \text{iff} \quad \quad \left[\forall v \in \mathcal{C}: \quad \text{if} \quad \left( \forall \gamma \in \Gamma: v(\gamma)=\textbf{t}\right) \quad \text{then} \quad v(a)=\textbf{t}\right].
\end{equation}
\end{definition}
Any $H$-class thus defines the same consequence relation as the class $\mathcal{C}^*_{H}$ induced by the atlas $\mathcal{A_H}$: trivially, $\mathcal{C}^*_{H}$ is one particular $H$-class. Let us now analyze some properties of collections of $H$-classes.

\subsection{Uniqueness}
First, consider the following question: for a given atlas $\mathcal{A_H}$, are all $H$-classes isomorphic, i.e. are they all equivalent to $\mathcal{C}^*_{H}$? In other words, does the collection of all $H$-classes have more than one member? Let us briefly depart from QL and delve into an analogous question that can be raised about classical logic (CL). One can define the CL-consequence relation $\vDash_{CL}$ via the class of valuations $\mathcal{C}^*_{CL}$ determined by the standard truth-tables.\footnote{The class $\mathcal{C}^*_{CL}$ is defined by the following properties: for all $v \in \mathcal{C}^*_{CL}$, $v(a \land b)=\textbf{t}$ if and only if $v(a)=v(b)=\textbf{t}$; $v(a \vee b)=\textbf{f}$ if and only if $v(a)=v(b)=\textbf{f}$; $v(\lnot a)=\textbf{t}/\textbf{f}$ if and only if $v(a)=\textbf{f}/\textbf{t}$.} Can the same consequence relation be characterized by a different class of valuations? Is the class $\mathcal{C}^*_{CL}$ unique in this sense? This question has been answered by Carnap (1943), where he showed that there are non-standard classes of valuations that characterize the same relation $\vDash_{CL}$. One such class is given by $\mathcal{C}^*_{CL} \cup \left\{\tilde{v}_{CL}\right\}$, where the non-standard valuation $\tilde{v}_{CL}$ obeys
\begin{equation}
    \tilde{v}_{CL}(a)=\textbf{t}, \quad \text{if and only if `a' is a classical tautology (i.e. $\vDash_{CL}a$)}.
\end{equation}
It is easy to see that $\mathcal{C}^*_{CL}$ and $\mathcal{C}^*_{CL} \cup \left\{\tilde{v}_{CL}\right\}$ define the same relation of logical consequence, despite them being non-isomorphic.\footnote{Carnap saw this as a problem because it implies that no axiomatic calculus for CL can be considered what he called a ``full formalization'' of CL, i.e., no such calculus can uniquely determine the intended class of valuations (i.e. the one given by the standard truth-tables). This arguably presents a problem for the logical inferentialist thesis that ``rules fix meaning''. For an overview of inferentialism, including this problem, see (Murzi and Steinberger 2017).} Carnap's answer can be trivially transferred to the case of QL-consequence relations in the following way. Let $\mathcal{C}^*_{H}$ be the class of valuations induced by a quantum atlas $\mathcal{A_H}$, and $\vDash_{H}$ its corresponding QL-consequence relation. Consider the valuation $\tilde{v}_{H}$, given by:
\begin{equation}
    \tilde{v}_{H}(a)=\textbf{t}, \quad \text{if and only if `a' is an $H$-tautology (i.e. $\vDash_{H}a$)}.
\end{equation}
Again, the two different classes, $\mathcal{C}^*_{H}$ and $\mathcal{C}^*_{H}\cup \left\{\tilde{v}_H\right\}$, determine the same relation $\vDash_{H}$, and are thus both $H$-classes. Hence, similarly to the case of CL, there is more than one valuation compatible with one and the same consequence relation. We deem this worth of being emphasized in the following
\begin{proposition}\label{prop: uniqueness}
For any quantum atlas $\mathcal{A_H}$, there are at least two $H$-classes.  
\end{proposition}

\subsection{Truth-functionality}
Now we will continue the analysis of $H$-classes by inspecting their truth-functionality or lack thereof. In order to do so, let us first introduce the following definition.
\begin{definition}
Let $\mathcal{C}$ be a class of valuations on a sentential language $\mathcal{L}$. We say that $\mathcal{C}$ makes `$\land$' truth-functional (TF) if there exists a function $f_{\land}$, such that for all $v \in \mathcal{C}$, and any $a,b \in S$:
\begin{equation*}
    v(a \land b)=f_{\land}(v(a),v(b)).
\end{equation*}
The respective definitions for `$\vee$' and `$\lnot$' then follow by analogy. Furthermore, we say that $\mathcal{C}$ is a truth-functional (TF) class of valuations if it makes all three connectives in $\left\{\land,\vee,\lnot\right\}$ truth-functional.
\end{definition}
Before applying this definition to QL, let us again mention a few facts about CL. Note first that the CL-class of valuations determined by the standard truth-tables, i.e., $\mathcal{C}^*_{CL}$, is obviously TF, and is actually the only TF CL-class. Nevertheless, there are also non-TF CL-classes, as exemplified by Carnap's non-standard class $\mathcal{C}^*_{CL}\cup \left\{\tilde{v}_{CL} \right\}$, which makes disjunction and negation non-TF. The non-truth-functionality of disjunction can for instance be seen from the following example. For arbitrary atomic sentences $a,b$: $\tilde{v}_{CL}(a) = \tilde{v}_{CL}(\neg a) = \tilde{v}_{CL}(b) = \textbf{f}$, and $\tilde{v}_{CL}(a \vee b) = \textbf{f}$, but $\tilde{v}_{CL}(a \vee \neg a) = \textbf{t}$. This already indicates that statements such as ``Classical logic is truth-functional'' ought to be approached carefully, and should be precisified as ``There exists a (unique) TF CL-class'', but acknowledging at the same time that ``There exist non-TF CL-classes''.

Let us now get back to QL and inspect the (non-)truth-functionality of $H$-classes. We will first show that for any quantum atlas $\mathcal{A_H}$, the induced class $\mathcal{C}^*_{H}$ makes conjunction TF, and negation and disjunction non-TF. The truth-functionality of conjunction follows immediately from the fact that, in any Hilbert space, a 1-dimensional subspace is contained in two other subspaces if and only if it is contained in their intersection. Since the conjunction of two sentences is interpreted as the intersection of the subspaces associated (under the same interpretation) to those sentences, this implies that a conjunction is true if and only if both its conjuncts are true, in accord with the standard classical truth-table. On the other hand, in order to illustrate the non-truth-functionality of disjunction and negation, consider the following example. Take three 1-dimensional subspaces $P_1,P_2,P \in C(H)$, where $H$ is an arbitrary Hilbert space. Suppose that $P_1,P_2,P$ respectively contain vectors $\psi_1$, $\psi_2$ and $\frac{1}{\sqrt{2}}\left(\psi_1+\psi_2\right)$, where $\psi_1$ and $\psi_2$ are orthogonal. Take an arbitrary sentence $a \in S$ and an interpretation $i \in I$, such that $i(a)=P_1$ and $i(\lnot a)=P_2$. It is easy to see that for the induced valuations $v^{(i)}_P$: $v^{(i)}_P(a)=v^{(i)}_P(\lnot a)=\textbf{f}$ but $v^{(i)}_P(a \vee \lnot a)=\textbf{t}$. On the other hand, if $P'$ is a 1-dimensional subspace containing a vector which does not lie in the span of $\psi_1$ and $\psi_2$, then $v^{(i)}_{P'}(a)=v^{(i)}_{P'}(\lnot a)=\textbf{f}$ and $v^{(i)}_{P'}(a \vee \lnot a)=\textbf{f}$. Thus, disjunction is not TF, in that some false disjuncts combine into a true disjunction, whereas others combine into a false disjunction; similarly, some false propositions, when negated, become true, whereas others stay false. Therefore, the $H$-classes of valuations induced by quantum atlases are not TF, for they make only conjunction TF. But can there be any TF $H$-classes? As proved by David Malament (2002), not only is the answer negative, but any $H$-class must make \textit{at least two} connectives non-TF. Here is his result translated into our terms:
\begin{proposition}\label{prop: malament}
Let $\mathcal{A_H}$ be the quantum atlas built on some arbitrary Hilbert space $H$. There is no $H$-class that makes more than one connective in $\left\{\land,\vee,\lnot\right\}$ truth-functional. Thus, there is no truth-functional $H$-class. 
\end{proposition}
Notice that the classes $\mathcal{C}^*_{H}$ induced by atlases $\mathcal{A_H}$ accordingly make only conjunction TF. Are there $H$-classes that make disjunction or negation TF? Surprisingly, the answer is no! It is enough to notice that for any Hilbert space $H$, the induced QL-consequence relation $\vDash_{H}$ validates the following arguments, which are the ``semantic versions'' of the $\land$-introduction and $\land$-elimination rules: for any $a,b \in S$, $a \land b \vDash_{H} a$, $a \land b \vDash_{H} b$, and $\left\{a,b\right\}\vDash_{H} a \land b$. It is easy to see that the validity of these three arguments implies that for any $H$-class, $a \land b$ is true if and only if both $a$ and $b$ are true, thus forcing conjunction to behave classically. Since, following Proposition \ref{prop: malament}, at most one connective can be TF, this implies that there is no $H$-class that makes either disjunction or negation TF! Let us state this clearly as follows:
\begin{proposition}\label{prop: conj is cl}
Let $\mathcal{A_H}$ be a quantum atlas built on some arbitrary Hilbert space $H$. Every $H$-class makes conjunction truth-functional, but negation and disjunction non-truth-functional.  
\end{proposition}
Where does this asymmetry between the QL-connectives come from? It suffices to take a look at Definition \ref{val ql} of $H$-classes and to notice that it invokes a metalinguistic quantification over all premises: it defines valid arguments as those whose conclusion is true if \textit{each premise} is also true. This definition formalizes the intuitive idea that valid arguments preserve truth, or that given that the premises are true, the conclusion must be true as well. Crucially, ``the premises are true'' is formalized as ``each premise is assigned truth-value `\textbf{t}'", which then, given the validity of the $\land$-introduction and $\land$-elimination arguments forces the QL-conjunction to behave classically. 

Now, one may either accept this conclusion, i.e., that in QL only conjunction can be TF (since it actually must be TF), or one may modify some of the above definitions so as to allow for the possibility of having a TF negation or disjunction. Let us briefly sketch one such modification. Instead of formalizing ``the premises are true'' as ``each premise is assigned truth-value `\textbf{t}'", we propose to formalize it as ``the conjunction of the premises is assigned truth-value `\textbf{t}'". We thereby introduce the notion of \textit{$H$-klass} as follows.

\begin{definition}\label{val ql2}
Let $\mathcal{L}$ and $S$ be as before. Suppose the consequence relation $\vDash_{H}$ on $S$ is induced, again just as before, by a quantum atlas $\mathcal{A_H}$, for some Hilbert space $H$. An $H$-klass $\mathcal{K}$ is any class of valuations on $\mathcal{L}$ that obeys the following. For all finite sets $\Gamma \cup \left\{a\right\} \subseteq S$:
\begin{equation}\label{valuations2}
    \Gamma \vDash_{H} a \quad \quad \text{iff} \quad \quad \left[\forall v \in \mathcal{K}: \quad \text{if} \quad v\left(\bigwedge_{\gamma \in \Gamma}\gamma\right)=\textbf{t} \quad \text{then} \quad v(a)=\textbf{t}\right].
\end{equation}
\end{definition}
The notion of \textit{$H$-klass} is thus different from the notion of \textit{$H$-class} introduced in Definition \ref{val ql}. Even though both are arguably reasonable formalizations of one and the same intuitive idea, i.e., that in valid arguments the truth of the premises guarantees the truth of the conclusion, they are nevertheless distinct, in that the collection of all $H$-classes is different from the collection of all $H$-klasses. Some potential drawbacks of the alternative notion of $H$-klass are that it is defined only for finite sets of premises, and that it cannot be straightforwardly transposed to logics defined on languages with no conjunction.\footnote{We are assuming throughout this paper that $\mathcal{L}$ is a finitary language, i.e., it allows for the formation of only finite conjunctions and disjunctions. The consequences of dropping this restriction and considering infinitary languages are deferred to future research.} Let us however provisionally sweep these problems under the rug and explore where the introduction of $H$-klasses leads us. First, notice that since conjunctions of premises are just a specific instance of single premises, the following holds:
\begin{proposition}\label{prop: class-klass}
For any quantum atlas $\mathcal{A_H}$, all $H$-classes are also $H$-klasses, but not viceversa.
\end{proposition}
This implies that the $H$-class $\mathcal{C}^*_{H}$ induced by an atlas $\mathcal{A_H}$ is an $H$-klass. Next, recall that the $\land$-introduction and $\land$-elimination arguments force the QL-conjunction to obey the standard classical truth-table in all viable $H$-classes. On the other hand, the $\land$-introduction argument $\left\{a,b\right\}\vDash_{H} a \land b$ does not impose any constraint whatsoever on viable $H$-klasses, because it is morphed into the trivial requirement that if $v(a \land b)=\textbf{t}$ then $v(a \land b)=\textbf{t}$. Therefore, while it is still the case that true conjunctions necessarily have true conjuncts (due to the validity of the $\land$-elimination argument), it may be the case that some false conjunctions have true conjuncts. Indeed, consider the following $H$-klass. For each subspace $P \in C(H) \setminus \left\{\mathbf{0},H \right\}$, define the function $h_P: C(H) \rightarrow \left\{\textbf{t},\textbf{f}\right\}$ as $h_P(Q)=\textbf{f}$ if and only if $Q \subseteq P$. Notice that $Q_1 \subseteq Q_2$ holds if and only if for all $P$, if $h_P(Q_2)=\textbf{f}$ then $h_P(Q_1)=\textbf{f}$, or contrapositively, if $h_P(Q_1)=\textbf{t}$ then $h_P(Q_2)=\textbf{t}$. This entails that the class $\mathcal{C}^{(\vee)}_{H}\equiv\left\{h_P \circ i, \quad \forall i \in I, \forall P \in C(H)\setminus \left\{\mathbf{0},H\right\} \right\}$ is an $H$-klass. In order to see that $\mathcal{C}^{(\vee)}_{H}$ makes disjunction TF notice the following. For any triple $P_1,P_2,Q \in C(H)$ it holds that $P_1 \subseteq Q$ and $P_2 \subseteq Q$ if and only if $P_1 \sqcup P_2 \subseteq Q$. According to $\mathcal{C}^{(\vee)}_{H}$, this implies that a sentence associated to $P_1$ and a sentence associated to $P_2$ are both false if and only if their disjunction is also false, thus mimicking the standard classical truth-table for disjunction. Since, as it can be easily checked, Proposition \ref{prop: malament} extends to $H$-klasses as well, $\mathcal{C}^{(\vee)}_{H}$ must make conjunction and negation non-TF: indeed, some true conjuncts form true conjunctions and others form false ones, whereas negation maps some true propositions into false ones and others into true ones. There is thus at least one $H$-klass that makes disjunction TF. Also, due to Proposition \ref{prop: class-klass}, there is at least one $H$-klass that makes conjunction TF, namely $\mathcal{C}^*_{H}$. Finally, a theorem proved in Friedman and Glymour (1972, 21) implies that for any Hilbert space $H$, there exists at least one $H$-klass that makes negation classical and thus TF, i.e., one for which `$a$' is true (false) if and only if `$\lnot a$' is false (true). Let us state these results as follows:
\begin{proposition}\label{prop: main ql}
Let $\mathcal{A_H}$ be a quantum atlas built on an arbitrary Hilbert space $H$. For any connective $c \in \left\{\land,\vee,\lnot\right\}$, there exists an $H$-klass that makes $c$ truth-functional.
\end{proposition}
Let us finally say a few words about similar issues in CL, since it will be relevant for our discussion in the subsequent section. Notice that, given the standard definition of CL-classes, the CL-validity of $\land$-introduction and $\land$-elimination rules imposes the standard truth-table on CL-conjunction as well, as opposed to disjunction and negation, for which Carnap's non-standard valuations exist\footnote{As noticed already by Carnap (1943) himself, one can, dually to the case of conjunction, impose the standard truth-table on CL-disjunction if one formalizes valid arguments as relations between sets of premises and \textit{sets of conclusions} and if one modifies the definition of CL-classes of valuations accordingly. The $\vee$-elimination and $\vee$-introduction rules, i.e. $a \vDash_{CL} a \vee b$, $b \vDash_{CL} a \vee b$ and $a \vee b \vDash_{CL} \left\{a,b\right\}$, then imply that a disjunction is false if and only if each of its disjuncts is false (see Shoesmith and Smiley 1978).} There are thus no CL-classes that make conjunction non-TF. However, analogously to the case of QL, if one defines \textit{CL-klasses} as those that make the conclusions of CL-valid arguments true if they make the \textit{conjunction of all the premises} true, then it turns out that there is at least one CL-klass that makes conjunction non-TF. Namely, the latter is given by $\mathcal{C}^*_{CL}\cup \left\{\tilde{v}'_{CL} \right\}$, where the non-standard valuation $\tilde{v}'_{CL}$ is defined as:
\begin{equation}
    \tilde{v}'_{CL}(a)=\textbf{f}, \quad \text{if and only if `a' is a CL-falsehood (i.e. $\vDash_{CL}\lnot a$)}.
\end{equation}
Since Carnap's non-normal CL-class makes disjunction and negation non-TF, and because, analogously to Proposition \ref{prop: class-klass}, every CL-class is also a CL-klass, we have obtained the following:
\begin{proposition}\label{prop: main cl}
For any connective $c \in \left\{\land,\vee,\lnot\right\}$, there exists a $CL$-klass that makes $c$ non-truth-functional.
\end{proposition}
Propositions \ref{prop: main ql} and \ref{prop: main cl} show that the truth-functional status of the QL- and CL-connectives is not as different as is often suggested: each connective can be TF under some klass of valuations and non-TF under some other klass of valuations. The truth-functionality of any of the three QL-connectives can thus be traded for the non-truth-functionality of the remaining QL-connectives. That said, the main difference between the QL- and CL-connectives still persists, in that - in accord with the extension of Proposition 2 to H-klasses - there is certainly no klass that makes \textit{all} the QL-connectives TF, which justifies to some extent the simplified claim that ``Quantum logic is not truth-functional''. Further below, in section five, we will be using Propositions \ref{prop: main ql} and \ref{prop: main cl} in our analysis of an argument that purports to show that CL- and QL-connectives differ in meaning. For now, let us however put to work the notions introduced above for the purpose of clarifying certain formal-semantic conditions for holding onto realism about quantum mechanics. 

\section{Semantics and Realism}
Our approach to QL has so far been rather abstract, in that we formulated QL-consequence relations on sets of sentences of an abstract sentential language. But the latter also admits of definitions of other logical consequence relations, such as the classical one, thereby enabling a more transparent comparison of the formal-semantic aspects of QL and CL, which will be useful in our discussion further below. We should emphasize, however, that our approach stands in contrast to the more common way of presenting QL as ``the logic of quantum mechanics'' (QM), which defines QL-consequence relations as relations between sentences that are intended to express ``experimental propositions'' about magnitudes associated to quantum systems/objects. In section two, we presented QL independently of QM, on a par with any other logic, similarly to how CL can obviously be conceived more abstractly than just as ``the logic of classical physics''. Nevertheless, it will now be useful to relate our presentation to the more common one, in order to assess the possible philosophical merit of the classes and klasses of valuations introduced in the previous section.

Following van Fraassen (1967), QL can be presented by first introducing a set of sentences $S_{QM}$ that is generated by conjoining, disjoining and negating ``elementary sentences'', each of which specifies the range of values taken by a magnitude associated to a quantum system/object. An elementary sentence is of the form ``sent($O$, $M$, $\Delta)$'', which reads  ``The value of magnitude $M$ associated to quantum object $O$ lies in the interval $\Delta$'', where $M$ is a magnitude, such as position or momentum, and $\Delta$ is some subset in the set of values that may possibly be taken by $M$, e.g., a subset of $\mathbb{R}$. One then introduces an interpretation function $h$ that maps each sentence in $S_{QM}$ to a closed subspace of the Hilbert space $H$ that represents the state-space of the quantum system under consideration. The map $h$ is of course not arbitrary, but is dictated by the empirical content of QM, in the following way. Each elementary sentence ``sent($O$, $M$, $\Delta$)'' is mapped to the closed subspace that contains all those states $\psi \in H$ for which it holds that, if $O$ is prepared in state $\psi$, the outcome of a measurement of $M$ on $O$ would lie in $\Delta$. Together with the latter constraint, the map $h$ acts on arbitrary sentences $a,b \in S_{QM}$ as follows:
\begin{equation}
    \begin{split}
        &h(a \land b)=h(a)\cap h(b)\\
        &h(a \vee b)=h(a) \sqcup h(b)\\
        &h(\lnot a)=h(a)^{\perp}.
    \end{split}
\end{equation}  
Finally, each quantum state $\psi \in H$ partially determines a potentially non-bivalent valuation $v_{\psi}$ on $S_{QM}$ as follows: for any sentence $a$, $v_{\psi}(a)=\textbf{t}$ if and only if $\psi \in h(a)$. The valuations $v_{\psi}$ may indeed be non-bivalent, as one does not need to equate ``\textit{a} is not true'' and ``\textit{a} is false''. We will, however, do so, for we will be interested in the question of realism in a moment.\footnote{The connection between bivalence and realism will be presently clarified.} In any case, given the above definition of valuations, a QL-consequence relation can be straightforwardly defined: a set of premises QL-implies a conclusion if for any $\psi \in H$, $v_{\psi}$ makes the conclusion true if it makes each premise true as well. 

Let us now relate all of this to our presentation of QL from section two above. Unlike the interpretation $h$ that assigns a fixed meaning to each sentence in $S_{QM}$, a quantum interpretationally constrained language $(\mathcal{L}, \mathcal{A_{H}}, I)$ contains all homomorphic interpretations $i \in I$ of the sentential language $\mathcal{L}$ in the quantum atlas $\mathcal{A_{H}}=(\mathcal{H},\langle D_P \rangle)$. Each designated set $D_P$ in turn contains those elements that are made true by the valuation $v_{\psi}$, for $\psi \in P$. The main difference between the two formulations is thus that ours leaves the association between sentences and propositions partially constrained, rather than completely fixed, thus ``erasing the intensional vestiges'' in the language of QL, as we already noted at the outset. It should be clear that if one takes the closed subspaces of $H$ to represent ``experimental propositions'' about a quantum system, and the elements of $H$ (or its 1-dimensional subspaces) to represent its possible quantum states, then $I$ contains all ways of compositionally assigning experimental propositions to the sentences of $\mathcal{L}$, whereas the induced class of valuations $\mathcal{C}^*_{H}$ contains ``$\psi$-relative'' valuations that make true all and only those propositions that can be verified with certainty in an appropriate measurement of the system prepared in state $\psi$. 

This finally brings us to discussing the philosophical importance of the formal results about valuations in QL presented in section three. It is well known that a realist understanding of QM -- that is, roughly, an understanding that views it as a theory that asserts truths about physical reality - runs easily into serious troubles and is still an open issue in contemporary science and philosophy. An important attempt at settling this issue was famously put forward by Putnam (1968), who deemed the adoption of QL as a necessary ingredient for maintaining a healthy realist view of the quantum world, devoid of unpalatable metaphysical hypotheses supposedly implied by theories that concord with CL (such as non-local hidden-variable theories). Putnam's bold proposal, which implied that all interpretational conundra of QM disappear upon acceptance of QL, and in particular that QL allows one to think of all observables in QM as having definite values at all times, was of course immediately put under serious scrutiny.\footnote{For example, Putnam's proposal was criticized among others by Kripke in a widely disseminated talk of 1974, recently published as (Kripke, 2024). For a critical review, see Stairs (2006, 2016).}

Importantly, in order to formally address the possible merit of Putnam's realism, Friedman and Glymour (1972) articulated the latter in a formal-semantic framework within which they assessed the possibility of providing the language of QL with a bivalent semantics. Bivalence can, indeed, be considered as a prerequisite for realism, at least in the present context, since it maintains that any fact expressed by a sentence either obtains or does not obtain in the physical reality, e.g., it is either the case or not the case that exactly $n$ hydrogen atoms are currently located in your room. However, as pointed out by Friedman and Glymour, Putnam's proposal is faced with immediate difficulties due to the Kochen-Specker (KS) theorem, which implies that, for Hilbert spaces of dimension $d>2$, there is no valuation that assigns definite values to all magnitudes, while respecting the functional relationships among the latter.\footnote{For some preliminary discussion of the consequences of the KS theorem for QL, see Dickson 1998, especially section 4.1.2. However, further clarification on the relationship between the non-truth-functionality of QL and the KS-theorem would be needed, but this is not required for our discussion in this paper.} Nevertheless, Friedman and Glymour made it clear that, while the KS-theorem does present problems for realism, at least as Putnam had conceived it, it certainly does not entail that the language of QM fails to admit a bivalent semantics.\footnote{In fact, as one referee pointed out, it is rather easy to see that any consequence relation admits of a bivalent semantics: for any valid argument $\left\{a_1, . . . ,a_n\right\} \centernot\vDash b$, the corresponding class of valuations is to contain valuation $v$ such that $v(a_1) = ... = v(a_n) = t$ and $v(b) = f$. That a bivalent semantics can be preserved for QL had been also expressed clearly by William Demopoulos: ``There are two different accounts of indeterminism which are historically important. The first, which apparently goes back to Aristotle, rejects bivalence: A theory is indeterministic if it assumes that there are propositions whose truth value is indeterminate. The second, represented by the quantum theory, retains bivalence while rejecting semi-simplicity [i.e., a property equivalent to truth-functionality]. ... This [latter] form of indeterminism implies that there is no Boolean representation of the properties obtaining at a given time; yet for any property \textit{P} it is completely determinate whether or not \textit{P} holds.'' (Demopoulos 1976, 76sq) Thus, it is one thing to say that it is true that ``This photon will decay tomorrow or this photon will not decay tomorrow'', while each of the disjuncts is neither true nor false, and another thing to say that it is true that ``This photon passed through the upper slit or this photon passed through the lower slit'', when each of the disjuncts is false. The conflation of the two different accounts noted by Demopoulos has unfortunately been repeatedly made in the literature (see, e.g., Bell and Hallett 1982, 368). Others made the same conflation on the basis of the Jauch-Piron theorem: ``Jauch and Piron show that any so-called orthomodular lattice (in particular any Hilbert lattice) admits total homomorphisms onto $\{0, 1\}$ iff it is distributive. Note that this means that any form of quantum logic \textit{must give up bivalence}" (Bacciagaluppi 2009, 56).} 

This conclusion also follows from the previous section, where we explicitly constructed a couple of bivalent semantics by defining appropriate classes and klasses of valuations. While none of these semantics can, due to the KS-theorem, provide in a straightforward way support to realism about QM, some of them may actually fare better than others in this respect. In fact, Friedman and Glymour, after presenting potential candidates for a realist bivalent semantics, immediately dismissed the ones that make negation non-classical, for ``dereliction of duty'' (1972, 20). They deemed it unacceptable for a realist semantics to ascribe the same truth-value both to a sentence and to its negation, thereby subscribing to the following principle.
\begin{principle}\label{principle neg}
A bivalent semantics is compatible with realism only if its negation obeys the classical truth-table (i.e. for any valuation $v$ and sentence `s': $v(s)\neq v(\lnot s)$).   
\end{principle}
One can of course raise doubts about this principle, especially in a context, such as the present one, in which classical logical notions are at stake. For example, it is not clear why one should endorse this principle without also endorsing parallel ones that would require any realist semantics to make conjunction and disjunction classical, thereby conflating realist and classical semantics.\footnote{Some ideas along these lines can be found in (Dummett 1976).} That is, why is it unpalatable for a realist to hold that both a sentence and its negation have the same truth-value, but simultaneously be at ease with false disjuncts making true disjunctions or true conjuncts making false conjunctions? What makes negation special? What makes its classicality an essential aspect of realism?

In any case, we are not going to further criticize Principle \ref{principle neg} in this paper, but we want to point out a formal difficulty posed by the results from the previous section. Namely, recall that, in accord with Proposition \ref{prop: conj is cl},  since any $H$-class makes only conjunction TF, it must necessarily make negation non-TF and thus non-classical. Thus, if one accepts Principle \ref{principle neg}, then no $H$-class can provide a semantics compatible with realism! Nevertheless, if we amend the formal treatment of arguments with multiple premises, as we have done by introducing the notion of \textit{H-klasses}, and since this amendment arguably does not threaten realism, then there are indeed bivalent semantics that can support a sort of realism about QM, albeit certainly not the sort of realism Putnam had wished for. In particular, there are bivalent semantics based on $H$-klasses that make negation TF, while making conjunction and disjunction non-TF.

Let us now take stock. We have argued that since QL can be considered as the logic of experimental propositions in QM, and bivalence is widely understood as a prerequisite for realism, formal-semantic results concerning bivalent semantics in QL are consequential for the possibility of maintaining realism in the quantum domain. Even though the KS-theorem presents difficulties for a ``naive realist'' understanding of QM, one can still salvage bivalence, and thereby a weaker form of realism, by giving up truth-functionality. Furthermore, even if one assumes that the truth-functionality of negation is a necessary requisite for a semantics to be compatible with realism, the latter might still be maintained if one were ready to amend the treatment of arguments with multiple premises. All in all, realism about QM and bivalence in QL are definitely not (yet) ruled out, though they are certainly rather costly from a semantic point of view.

\section{Beyond Realism: Meaning Variance}
As seen in the previous section, Putnam's claim that QL can be used to furnish a realist picture of the quantum world has been shown to require a serious revision, mainly due to the no-go result by Kochen and Specker.\footnote{To be sure, Putnam himself admitted as much. Furthermore, already in papers from the early 1990s, he acknowledged that a realist understanding of QM need not embrace QL. See, e.g., Putnam 1991 and Putnam 1994. Also, for a later acknowledgement, see Putnam 2012.} There is however a further bold proposal made in the same 1968 paper, in that Putnam argued that QL is the ``one true logic'' and that the world we live in is a ``quantum logical world''. Moreover, he considered QL not merely as a replacement of CL that may play a role in reasoning about quantum objects, but as a revision of CL, that is, an alternative logic that asserts different truths about the same connectives. More specifically, he maintained that ``adopting quantum logic is \textit{not} changing the meaning of the logical connectives, but merely changing our minds about the [distributive] law.'' (Putnam 1968, 233) In other words, QM did not just lead us to the introduction of a new mathematical structure that we may interpret as a logic, but to the discovery that the ``one true logic'' is non-classical.\footnote{Of course, others have denied precisely the claim that this new mathematical structure \textit{should} be interpreted as a logic, for the relation $\vDash$ is not really a relation of logical consequence, and the lattice operations $\land, \lor, \neg$ are not really logical connectives. This debate is, however, orthogonal on the line of argumentation that we, following Hellman, pursue in this paper.} This claim sparked another line of criticisms (e.g., in Fine 1972, Dummett 1976), which do not challenge primarily Putnam's realism, but his logical revisionism; they contend, more particularly, that QL cannot be considered a revision of CL, since the two are incommensurable:  their connectives do not mean the same thing. 

A criticism of this latter kind was prominently put forward by Geoffrey Hellman, who commented: ``[T]he opponent against whom Putnam argued was a rather dogmatic conventionalist who was rather prone to put more weight on the notion of `meaning' than scientific scrutiny should allow. What I want to do here is focus on a more precise `meaning-change' argument, one which makes absolutely minimal reliance on the problematic word, `meaning', and which, as far as I can see, a proponent of Putnam's view can neither defeat nor bypass.'' (Hellman 1980, 494) In what follows, we reconstruct Hellman's argument and show how it can, and should, be improved in light of the discussion in the previous sections. 

Hellman started by stipulating a condition for meaning invariance, which would presumably be acceptable to a conventionalist without dogmas: 
\begin{principle}\label{principle hellman}
``If $\alpha$ and $\beta$ are synonymous sentential connectives, then (a) if one is a truth-functional connective, then so is the other, and (b) if $\alpha$ and $\beta$ are truth-functional, they have the same truth tables.'' (Hellman 1980, 495)
\end{principle}
He went on to prove that \textit{if} QL-negation is TF, \textit{then} QL-disjunction and QL-conjunction are non-TF, which, due to clause (a) in Principle \ref{principle hellman}, implies that at least \textit{some} QL-connectives and their classical counterparts are not synonymous. The argument can be reconstructed as follows:

\bigskip

1. Two connectives have the same meaning only if they are either both truth-functional or both non-truth-functional.

\smallskip

2. If QL-negation is truth-functional, then QL-disjunction and QL-conjunction are not truth-functional.

\smallskip

3. Thus, QL-negation and CL-negation differ in meaning, or CL-disjunction (conjunction) and QL-disjunction (conjunction) differ in meaning.

\smallskip

4. Thus, some QL connectives differ in meaning from their CL counterparts.

\bigskip

Importantly, besides having the benefit of relying minimally on the problematically vague notion of `meaning', Hellman maintained that his argument is independent of the issues concerning realism, in that it is based on a purely formal-semantic result that exhibits the difference in the truth-functional status of the classical and quantum connectives: ``the non-truth-functionality argument is entirely distinct from [those that] argued that QL could not satisfy the demands of realism.'' (1980, 496) Now, notice that while the argument does establish that at least some QL connectives differ in meaning from their CL counterparts, it does not specify which ones do so unconditionally. Moreover, Conclusion 3 is compatible with QL-negation being non-TF and thus differing in meaning from CL-negation, while leaving QL-conjunction and QL-disjunction TF and thus possibly synonymous with their CL counterparts. Consequently, since the distributive law refers to conjunction and disjunction, but not to negation, Hellman's argument does not conflict with Putnam's logically revisionist claim that classical distribution fails in QM. In other words, despite Hellman's hope, the argument does not do enough to justify Fine's earlier claim that ``the sense of the distributive law in which it is said to fail is not the sense in which, as the distributive law, it is supposed to hold.” (Fine 1972, 19). More needs to be done for this to be the case. Our goal in the remainder of the paper is to use the results from section three in order to improve on the above argument. The question whether the improved argument definitively justifies Fine's claim will be touched upon at the end of our discussion.

First, notice that the conditional statement expressed in Premise 2 follows from Proposition \ref{prop: malament}, which states that any viable class of valuations can make at most one of the sentential connectives TF. Now, while the antecedent of Premise 2, stating that negation is TF, can arguably be justified for semantics that are intended to support realism - as required by Principle \ref{principle neg} - this assumption is not warranted in the present discussion, in which we, together with Hellman, are aiming to contrast CL- and QL-connectives on non-metaphysical, semantic grounds. Moreover, not only is the aforementioned antecedent not warranted, but it cannot be maintained without invalidating the quantum-logically valid $\land$-introduction argument, $\left\{a,b\right\} \vDash a \land b$. Indeed, recall that Proposition \ref{prop: conj is cl} states that any viable $H$-class of valuations makes conjunction TF, but disjunction and negation non-TF. Therefore, the antecedent of Premise 2, whose truth is needed if the argument is to back Fine's claim against Putnam's, is not only unwarranted, but also false\footnote{Note that this point is distinct from the often raised criticism that the quantum logician's metalanguage should obey quantum logic, in which case Hellman (1980)'s proof for the non-truth-functionality of conjunction and disjunction would not go through.}.

Furthermore, there is another qualification that needs to be made here: namely, it is not entirely clear what is meant in the above argument by saying that a sentential connective `c' is TF. Does it mean that (i) \textit{there exists} a viable class of valuations that makes `c' TF, or that (ii) \textit{any} viable class of valuations makes `c' TF? Since the argument tacitly assumes that classical connectives are TF, option (ii) is automatically excluded, since the existence of Carnap's non-standard class $\mathcal{C}^*_{CL}\cup \left\{\tilde{v}_{CL} \right\}$ would make even CL-disjunction and CL-negation non-TF! Therefore, the only chance for a Hellman-type argument to establish meaning-variance on the basis of a difference in the truth-functional status of CL- and QL-connectives is to stick to option (i). 

With all this in mind, we now offer an improved argument, which will, importantly, still rely on the same condition for meaning invariance stipulated by Hellman, i.e., Principle \ref{principle hellman}.
Let us say that a sentential connective `c' is TF if there is at least one viable class of valuations that makes `c' TF. Here is our argument:

\bigskip

1'. Two connectives have the same meaning only if they are either both truth-functional or both non-truth-functional.

\smallskip

2'. QL-disjunction and QL-negation are not truth-functional.

\smallskip

3'. Thus, CL-disjunction (negation) and QL-disjunction (negation) differ in meaning.

\smallskip

4'. Thus, some QL connectives differ in meaning from their CL counterparts.

\bigskip

This argument can be understood as a completion of Hellman's, in that, while it maintains the same general conclusion (i.e. the one expressed in 4'.), it also specifies which QL-connectives are not synonymous with their classical counterparts: in particular, since disjunction - which figures in the distributive law - is among these connectives, our argument appears, at least on the surface, to justify Fine's claim.

There is, however, another point we want to make concerning the above argumentation. Namely, notice that we have so far assumed that the only viable classes of valuations for QL are those introduced by Definition \ref{val ql}, i.e., $H$-classes. But recall that, after observing that both CL- and QL-conjunction are necessarily made TF by any viable $H$-class -- essentially due to the particular formal explication of the validity of arguments with multiple premises -- we introduced (in Definition \ref{val ql2}) the notion of $H$-klasses of valuations, thereby restoring the semantic symmetry between conjunction and disjunction. Now, how might considering $H$-klasses in the present discussion change anything? It can easily be seen that this is a problem both for Hellman's argument and for our own as well. For recall that, after introducing $H$-klasses, we went on to show that for any QL-connective, there is a klass that makes it TF, as stated by Proposition \ref{prop: main ql}, and that for any CL-connective, there is a klass that makes it non-TF, as stated by Proposition \ref{prop: main cl}. It thus follows that there is no difference anymore in the truth-functional status of CL- and QL- connectives, at least at the level of individual connectives: for any connective `c', there exists a klass that makes CL-`c' TF and there exists a klass that makes QL-`c' TF. Also, there exists a klass that makes CL-`c' non-TF and there exists a klass that makes QL-`c' non-TF. Therefore, if one formalizes the relation between truth and logical consequence by means of klasses, no meaning-variance can be established by any Hellman-type argument (i.e. one that relies on Principle \ref{principle hellman}).
On the other hand, as already emphasized before, there is of course still a big semantic difference between CL and QL at the global level, in that, unlike in the case of CL, there is no truth-functional $H$-klass, i.e., no klass that makes \textit{all} three sentential QL-connectives TF. But this difference cannot be used to support a Hellman-type argument, since Principle \ref{principle hellman} considers only individual connectives and their counterparts.

To summarize, the soundness of a Hellman-type argument that implies meaning-variance between classical and quantum connectives depends on the formal explication of valid arguments with multiple premises. If one chooses to explicate such arguments in terms of \textit{klasses} of valuations, then any Hellman-type argument turns out to be unsound, as we have just seen. Alternatively, if one opts for \textit{classes} of valuations, then a sound argument that establishes meaning-variance can indeed be given (again, provided that one accepts Principle \ref{principle hellman}).
This is the argument we have presented above.

Importantly, however, for our argument to ultimately justify Fine's claim - that the quantum logician's rejection of distributivity does not conflict with the classical logician's assertion thereof - one would still need to argue (i) that classes, rather than klasses, are indeed the right way to go in formalizing the relationship between truth and validity, and (ii) that we should accept Principle \ref{principle hellman} as reliably tracking relations of synonymy between logical connectives. While this would demand a much broader investigation, here we want to offer a few preliminary thoughts on both points.

Let us start with (i). To the best of our knowledge, we are the first to have introduced the distinction between classes and klasses of valuations, and to have noticed that the latter restore the truth-functional symmetry between conjunction and disjunction (both in CL and in QL). The reason this distinction has been so far overlooked is presumably the widely held agreement that conjunction and truth are to be related to each other \textit{classically}, i.e. that a conjunction is true if and only if both of its conjuncts are true. This, of course, collapses the difference between classes and klasses, for their definitions then turn out to be equivalent. However, the distinction plays an important role in our \textit{formal-semantic} analysis above, which does not reject by fiat non-standard interpretations of logical connectives, such as Carnap's interpretation (which allows for true disjunctions with false disjuncts) and our own interpretation (which, dually, allows for false conjunctions with true conjuncts). If one finds these interpretations bizarre, one should be reminded that conjunction and disjunction -- even interpreted in these unusual ways -- still obey their usual QL- or CL-introduction and elimination rules. Thus, the distinction between classes and klasses need not collapse, and it does not collapse in the context of our formal-semantic analysis. Furthermore, we can see no serious reason why klasses of valuations, where the object-level conjunction is used for the formalization of arguments with multiple premises, should be considered inherently pathological. But whether this is so, or whether classes are instead the right way -- or at least the better way -- to formalize such arguments, cannot be decided here.

Proceeding with (ii), it is clear enough that both in Hellman's argument and in ours, Principle \ref{principle hellman} does the philosophical heavy-lifting, in that it proposes a connection between the somewhat loose and evasive notion of ``sameness of meaning'' (or ``synonymy'') and the formal-semantic notion of truth-functionality. As Hellman hoped, his principle would place that loose and evasive notion under transparent scientific-mathematical scrutiny. However, one should seek to justify the principle, lest it remain an arbitrary stipulation. Moreover, its justification ought to deal directly, rather than via formal-semantic proxies, with the looseness and evasiveness that Hellman's own appeal to the principle tried to avoid. Let us briefly illustrate one serious worry that comes to mind, and shows that the issue here cannot be resolved without doing at least \textit{some} metaphysics, and therefore, at least \textit{some} difficult not-merely-formal-semantic analysis.

Principle \ref{principle hellman} speaks of the relation of synonymy holding, or not holding, between logical connectives that are parts of certain logical theories (e.g., CL and QL). It is, however, not clear what the ``meaning'' of a connective is constrained by, and in particular, how it is related to the formal-semantic characterization that the theory ascribes to that connective. For even after acknowledging all of the formal-semantic differences between their theories, the classical and the quantum logician might still maintain that they are aiming to theorize about the logical relations that \textit{really} hold between propositions. They might claim, for example, that in using their respective negations and disjunctions, each of them aims to say something about negation and disjunction \textit{themselves}. In this case, the complaint that the quantum logician just cannot mean the same thing as the classical logician when uttering `not' and `or' may be rejoined by insisting that the former is actually correcting the latter's `not' and `or', and by further pointing out that there is no good reason why meaning should be constrained by formal-semantic properties. Note, however, that this appeal to the logical connectives themselves, as they really are, prior to any logician's attempt to theorize about them, betrays a ``logical-realist'' commitment, which maintains that something akin to a logical structure is inherently present in something akin to a world, e.g., our natural world, or the world of Fregean thoughts-propositions (McSweeney 2019). Indeed, we suspect that someone with logical-realist inclinations - such as Putnam, with his appeal to the ``one true logic'' - might have reasons to flatly reject Principle \ref{principle hellman}. However, discussing whether some form of logical-realism is to be maintained at all, or whether some alternative is to be preferred, and additionally, whether this alternative can make sense of the principle under discussion, would push us into murky philosophical waters spilling far beyond our formal-semantic analysis in this paper.

Wrapping up, we believe that our argument represents an improvement on Hellman's, and we find the improvement significant, for \textit{within the scope of the same assumptions} it provides a better justification for Fine's claim against Putnam. Nevertheless, we hesitate to declare that we have definitively vindicated that claim. The ultimate vindication would need, at the very least, to deal more fully with those very assumptions, i.e., with the problems sketched above concerning (i) the relationship between formal and pre-formal arguments with multiple premises, and, arguably more importantly, with (ii) the metaphysics of logic. Neither of these can reasonably be addressed in this paper.


\section{Conclusion}

In this paper, we started by presenting QL in the abstract-algebraic framework presented in (Dunn and Hardegree 2001), thus considering the language of QL as an interpretationally constrained language. Within this framework, we then analyzed the truth-functional status of QL-connectives, and introduced a distinction between two different types of classes of valuations: $H$-classes, which make conjunction truth-functional, and $H$-klasses, which allow for a non-truth-functional conjunction. Turning to more philosophical concerns, we first pointed out that a bivalent semantics of QL, that could potentially provide a basis for realism, is not ruled out. However, the price that must be paid by the realist is the truth-functionality of the QL-connectives. This led us to discuss Hellman's anti-revisionist argument that purports to show, on the basis of the aforementioned lack of truth-functionality, that adopting QL changes the meaning of some logical constants. We provisionally embraced his condition for meaning invariance, but pointed out that the argument does not really support a rejection of Putnam's revisionism. In fact, the argument would provide such a rejection only if QL-negation were truth-functional, which however, besides being unjustified, also turns out to be false, as it conflicts with the quantum-logically valid $\land$-introduction argument. One may in response want to consider $H$-klasses of valuations, but this only causes more trouble. In contrast, our modified argument fares better and makes the point that adopting QL changes the meaning of negation and disjunction. However, it does so only by taking on board the same assumptions that Hellman originally took: both Principle \ref{principle hellman} and the previously unnoticed choice of classes over klasses. We ended our discussion by briefly indicating the difficulties that a reflection on these assumptions would need to face - difficulties arguably shared by any attempt at mounting an anti-logical-revisionist argument based on merely formal-semantic considerations. Despite not reaching a definite verdict on Fine's anti-revisionist claim, we hope that the results developed throughout the paper - the abstract algebraic formulation of QL, the distinction between classes and klasses of valuations, and the comparison of the truth-functional status of CL and QL - would nevertheless be useful in addressing other problems in philosophy of logic.


\section*{Acknowledgments}
We thank David Malament for establishing the result we use in this paper and for his kind encouragement to pursue ours; an anonymous reviewer at this journal and another anonymous reviewer at another journal for constructive criticism that helped improve the paper; and audiences at the University of Vienna, Central European University in Vienna, and the Institut Henri Poincaré in Paris for useful discussion. The Austrian Science Fund (FWF) is gratefully acknowledged for financial support through Grants 10.55776/DOC162 and 10.55776/P36994 (for S.H.) and 10.55776/PAT3440123 (for I.D.T.). 

\section*{References}

Bacciagaluppi, G. (2009). Is Logic Empirical? In Gabbay D., D. Lehmann, and K. Engesser (eds.) \textit{Handbook of Quantum Logic}, Elsevier, Amsterdam, 49-78. 



\smallskip

Birkhoff, G. and J. von Neumann (1936). The Logic of Quantum Mechanics. In \textit{Annals of Mathematics}, 37, 823-843.

\smallskip

Carnap, R. (1943). \textit{Formalization of Logic}, Harvard University Press.

\smallskip

Demopoulos, W. (1976). The Possibility Structures of Physical Systems. In W. L. Harper and C. A. Hooker (eds.) \textit{Foundations of Probability Theory, Statistical Inference, and Statistical Theories of Science}, Dordrecht, Reidel, 55-80.



\smallskip

Dickson, M (1998). \textit{Quantum chance and non-locality}, Cambridge University Press.

\smallskip

Dummett, M. (1976). Is Logic Empirical? In \textit{Truth and Other Enigmas}, London, Duckworth, 269-289.

\smallskip

Dunn, J. M., \& Hardegree, G. (2001). \textit{Algebraic methods in philosophical logic}, Oxford University Press.

\smallskip

Fine, A. (1972). Some Conceptual Problems with Quantum Theory. In Robert G. Colodny (ed.) \textit{Paradigms and Paradoxes: The Philosophical Challenge of the Quantum Domain}, University of Pittsburgh Press,  3-31. 

\smallskip



Hellman, G. (1980). Quantum Logic and Meaning. \textit{Proceedings of the Philosophy of Science Association}, 493-511.

\smallskip

Kochen, S. and E. P. Specker (1967). The Problem of Hidden Variables in Quantum Mechanics. In C. A. Hooker (ed.) \textit{The Logico-Algebraic Approach to Quantum Mechanics}, I, Dordrecht, Reidel, 1975, 293-328.

\smallskip

Kripke, S. A. (2024). The question of logic. \textit{Mind, 133}(529), 1-36.

\small

Malament, D. B. (2002). Notes on Quantum Logic. Unpublished lecture notes.

\smallskip

McSweeney, M. M. (2019). Logical realism and the metaphysics of logic. \textit{Philosophy Compass,
14} (1), e12563

\smallskip

Murzi, J. and F. Steinberger (2017). Inferentialism. In B. Hale \textit{et al}. (eds.), \textit{A companion to the philosophy of language}, Blackwell, 197-224. 

\smallskip



Putnam, H. (1968). Is Logic Empirical? In R. S. Cohen and M. W. Wartofsky (eds.) \textit{Boston Studies in the Philosophy of Science}, 5, Dordrecht, Reidel, 216-241, reprinted as ``The logic of quantum mechanics'', in \textit{Mathematics, Matter and Method. Philosophical Papers}, 1, Cambridge University Press, 174-197.

\smallskip

Putnam, H. (1991). Il principio di indeterminazione e il progresso scientifico. \textit{Iride}, 7, 9-27.

\smallskip

Putnam, H. (1994). Michael Redhead on quantum logic. In P. Clark and B. Hale (eds.) \textit{Reading Putnam}, Blackwell, 265-280.

\smallskip

Putnam, H. (2012). The Curious Story of Quantum Logic. In M. De Caro and
D. Macarthur (eds.) \textit{Philosophy in the Age of Science: Physics, Mathematics, and Skepticism}, Harvard University Press, 162--177.

\smallskip

R\'edei, M. (1998). \textit{Quantum Logic in Algebraic Approach}, Dordrecht, Kluwer Academic Publishers.





\smallskip

Shoesmith, D. J. and T. Smiley (1978). \textit{Multiple Conclusion Logic}, Cambridge University Press.

\smallskip

Stairs, A. (2006). Kriske, Tupman and Quantum Logic: The Quantum Logician's Conundrum. In \textit{Physical Theory and its Interpretation: Essays in Honor of Jeffrey Bub} (pp. 253-272). Dordrecht: Springer Netherlands.

\smallskip

Stairs, A. (2016). Could logic be empirical? The Putnam-Kripke debate. In \textit{Logic and algebraic structures in quantum computing}, 45, 23.

\smallskip

Stone, Marshall H. (1936). The Theory of Representations of Boolean Algebras. \textit{Transactions of the American Mathematical Society}, 40, 37-111.



\smallskip

van Fraassen, B. C. (1967). Meaning relations among predicates. \textit{Noûs}, 161-179.



\end{document}